\documentstyle[aps,12pt]{revtex}
\begin{document}
\title{Relativistic Hartree-Bogoliubov description\\
of the deformed N = 28 region}
\author{ G.A. Lalazissis$^{1}$, D. Vretenar$^{1,2}$, P. Ring$^{1}$,  
M. Stoitsov$^{1,3}$, and L. Robledo$^{4}$} 
\address{
$^{1}$ Physik-Department der Technischen Universit\"at M\"unchen,
D-85748 Garching, Germany\\
$^{2}$ Physics Department, Faculty of Science, University of
Zagreb, Croatia\\
$^{3}$Institute for Nuclear Research and Nuclear Energy, \\
Bulgarian Academy of Sciences, 
Sofia--1784, Bulgaria\\
$^{4}$ Departamento de F\' isica Te\' orica C-XI, Universidad
Aut\' onoma de Madrid, E-28049, Spain\\ }
\maketitle
\bigskip
\bigskip
\begin{abstract}
Ground-state properties of neutron-rich $N \approx 28$ nuclei
are described in the framework
of Relativistic Hartree Bogoliubov (RHB) theory. The model uses
the NL3 effective interaction in the mean-field Lagrangian,
and describes pairing correlations
by the pairing part of the finite range Gogny interaction D1S.
Two-neutron separation energies and ground-state quadrupole
deformations that result from fully self-consistent
RHB solutions, are compared with available experimental data.
The model predicts a strong suppression of the spherical $N=28$ shell
gap for neutron-rich nuclei: the $1f7/2 \rightarrow fp$ core 
breaking results in deformed ground states.
Shape coexistence is expected for neutron-rich Si, S and Ar isotopes.
\end{abstract}
\bigskip
\section {Relativistic Hartree-Bogoliubov theory with
finite range pairing interaction}

Ground states of deformed exotic nuclei, unstable isotopes with
extreme isospin values, display many interesting and unique
properties. The description of observed phenomena, as well as 
the prediction of new and unexpected properties, present an exciting 
challenge for modern theoretical advances. For neutron-rich nuclei,
a fascinating example of the modification of the effective single-nucleon
potential is the observed suppression of shell effects, the disappearance
of spherical magic numbers, and the resulting onset of deformation 
and shape coexistence. Fine examples are $^{32}$Mg (extreme quadrupole
deformation) and $^{44}$S (shape coexistence). Isovector quadrupole 
deformations are expected at the neutron drip-lines, and possible 
low-energy collective isovector modes have been predicted. 
A very interesting phenomenon is a possible formation of deformed 
halo structures in weakly bound nuclei. In much heavier neutron-rich 
systems, the modification of shell structure could produce an enhancement
of stability for superheavy elements $Z\geq 110$. Proton-rich nuclei
present the opportunity to study the structure of systems beyond the 
drip-line. The phenomenon of ground-state proton radioactivity
is determined by a delicate interplay between the Coulomb and
centrifugal terms of the effective potential. Proton decay rates indicate 
that the region $57\leq Z \leq 65$ contains strongly deformed 
nuclei at the drip-lines. The lifetimes of deformed proton emitters
provide direct information on the shape of the nucleus.

And while an impressive amount of experimental data on deformed exotic 
nuclei has been published recently, relatively few pertinent theoretical
studies have been reported. For relatively light exotic nuclei, as
for example $^{32}$Mg, shell model predictions for the onset of 
deformation and shape coexistence have been remarkably successful.
However for medium-heavy and heavy systems, the only viable approach at 
present are large scale self-consistent mean-field calculations
(Hartree-Fock, Hartree-Fock-Bogoliubov, relativistic mean-field).
The problem is of course that, in addition to the self-consistent mean-field
potential, pairing correlations have to be included in order to
describe ground-state properties of open-shell nuclei. And while 
for strongly bound systems pairing can be included in the 
simple BCS scheme in the valence shell,  
exotic nuclei with extreme isospin values require a careful treatment 
of the asymptotic part of the nucleonic densities, and therefore 
a unified description of mean-field and pairing correlations.
The nonrelativistic Hartree-Fock-Bogoliubov (HFB) and 
relativistic Hartree-Bogoliubov (RHB) models have been very successfully 
applied in the description of a variety of nuclear structure phenomena
in spherical exotic nuclei on both sides of the valley of $\beta$-
stability. For deformed exotic nuclei however, so far pairing has been only 
described in the BCS approximation. The BCS scheme presents only a
poor approximation for nuclei with small separation energies, i.e.
with the Fermi level close to the particle continuum. The inclusion
of deformation in HFB or RHB models with finite range pairing still
presents considerable difficulties, and it has not been possible 
yet to obtain reliable results in coordinate space calculations. In the present
work we report on the first application of the relativistic Hartree-Bogoliubov
theory with finite range pairing interaction to the structure
of deformed nuclei in the N=28 region of neutron-rich nuclei.

In the framework of the
relativistic Hartree-Bogoliubov model, 
the ground state of a nucleus $\vert \Phi >$ is represented 
by the product of independent single-quasiparticle states.
These states are eigenvectors of the
generalized single-nucleon Hamiltonian which
contains two average potentials: the self-consistent mean-field
$\hat\Gamma$ which encloses all the long range particle-hole ({\it ph})
correlations, and a pairing field $\hat\Delta$ which sums
up the particle-particle ({\it pp}) correlations. The 
single-quasiparticle equations result from the variation of the 
energy functional with respect to the hermitian density matrix $\rho$
and the antisymmetric pairing tensor $\kappa$. 
The relativistic model was formulated
in Ref.~\cite{KR.91}. In the Hartree approximation for
the self-consistent mean field, the relativistic
Hartree-Bogoliubov (RHB) equations read
\begin{eqnarray}
\label{equ.2.2}
\left( \matrix{ \hat h_D -m- \lambda & \hat\Delta \cr
                -\hat\Delta^* & -\hat h_D + m +\lambda} \right) 
\left( \matrix{ U_k({\bf r}) \cr V_k({\bf r}) } \right) =
E_k\left( \matrix{ U_k({\bf r}) \cr V_k({\bf r}) } \right).
\end{eqnarray}
where $m$ is the nucleon mass and $\hat h_D$ is the single-nucleon Dirac
Hamiltonian ~\cite{Rin.96}
\begin{equation}
\hat{h}_{D}=-i{\mathbf{\alpha \cdot \nabla }}+\beta (m+g_{\sigma }\sigma
({\mathbf r}))+g_{\omega }\tau _{3}\omega ^{0}({\mathbf r})+
g_{\rho }\rho ^{0}({\mathbf r})+
e{\frac{{(1-\tau _{3})}}{2}}A^{0}({\mathbf r}).  \label{dirh}
\end{equation}
The chemical potential $\lambda$  has to be determined by
the particle number subsidiary condition in order that the
expectation value of the particle number operator
in the ground state equals the number of nucleons. The column
vectors denote the quasi-particle spinors and $E_k$
are the quasi-particle energies. The Dirac Hamiltonian 
contains the mean-field potentials of the isoscalar 
scalar $\sigma$-meson, the isoscalar vector $\omega$-meson,
the isovector vector $\rho$-meson, as well as the 
electrostatic potential. 
The RHB equations have to be solved self-consistently with
potentials determined in the mean-field approximation from
solutions of Klein-Gordon equations.
The equation for the sigma meson contains the 
non-linear $\sigma$ self-interaction terms.
Because of charge conservation only the
third component of the isovector $\rho$-meson contributes. The
source terms for the Klein-Gordon equations are calculated 
in the {\it no-sea} approximation. In the present version of the 
model we do not perform angular momentum or particle number
projection.

The pairing field $\hat\Delta $ in (\ref{equ.2.2}) is an integral
operator with the kernel 
\begin{equation}
\label{equ.2.5}
\Delta_{ab} ({\bf r}, {\bf r}') = {1\over 2}\sum\limits_{c,d}
V_{abcd}({\bf r},{\bf r}') {\bf\kappa}_{cd}({\bf r},{\bf r}'),
\end{equation}
where $a,b,c,d$ denote quantum numbers
that specify the Dirac indices of the spinors, 
$V_{abcd}({\bf r},{\bf r}')$ are matrix elements of a
general relativistic two-body pairing interaction, and the pairing
tensor is defined 
\begin{equation}
{\bf\kappa}_{cd}({\bf r},{\bf r}') = 
\sum_{E_k>0} U_{ck}^*({\bf r})V_{dk}({\bf r}').
\end{equation}
In the applications of the RHB theory to spherical nuclei  
~\cite{Gonz.96,PVL.97,LVP.98,LVR.97,VPLR.98,LVR.98,VLR.98}
we have used a phenomenological non relativistic interaction in the pairing 
channel, the pairing part of the Gogny force 
\begin{equation}
V^{pp}(1,2)~=~\sum_{i=1,2}
e^{-(( {\bf r}_1- {\bf r}_2)
/ {\mu_i} )^2}\,
(W_i~+~B_i P^\sigma 
-H_i P^\tau -
M_i P^\sigma P^\tau),
\end{equation}
with the set D1S \cite{BGG.84} for the parameters 
$\mu_i$, $W_i$, $B_i$, $H_i$ and $M_i$ $(i=1,2)$.  
This force has been very carefully adjusted to the pairing 
properties of finite nuclei all over the periodic table. 
In particular, the basic advantage of the Gogny force
is the finite range, which automatically guarantees a proper
cut-off in momentum space. The fact that it is a non-relativistic
interaction has negligible influence on the results of RHB
calculations. Of course, in order to have a consistent formulation
of the model, the matrix elements in the pairing channel $V_{abcd}$ 
should be derived as a one-meson exchange interaction by eliminating
the mesonic degrees of freedom in the model Lagrangian~\cite{KR.91}.
However, although recently results have been reported in applications 
to nuclear matter, a microscopic and fully relativistic  
pairing force, derived starting from the Lagrangian of quantum
hadrodynamics, still cannot be applied to nuclei.
Therefore also in the present calculations we use the effective
Gogny interaction in the pairing channel.

The eigensolutions of Eq. (\ref{equ.2.2}) form a set of
orthogonal and normalized single quasi-particle states. The corresponding
eigenvalues are the single quasi-particle energies.
The self-consistent iteration procedure is performed
in the basis of quasi-particle states. The resulting quasi-particle
eigenspectrum is then transformed into the canonical basis of single-particle
states, in which the RHB ground-state takes the  
BCS form. The transformation determines the energies
and occupation probabilities of the canonical states.

The self-consistent solution of the 
Dirac-Hartree-Bogoliubov integro-differential eigenvalue equations
and Klein-Gordon equations for the meson fields determines the 
nuclear ground state. For systems with spherical symmetry, i.e. 
single closed-shell nuclei, the coupled system of equations has been
solved using finite element methods in coordinate space 
\cite{PVL.97,LVP.98,LVR.97,VPLR.98}, and by expansion in a basis
of spherical harmonic oscillator~\cite{Gonz.96,LVR.98,VLR.98}. For deformed
nuclei the present version of the model does not include solutions 
in coordinate space. 
The Dirac-Hartree-Bogoliubov equations and the equations for the 
meson fields are solved by expanding the nucleon spinors 
$U_k({\bf r})$ and $V_k({\bf r})$, 
and the meson fields in terms of the eigenfunctions of a 
deformed axially symmetric oscillator potential~\cite{GRT.90}. 
In the present calculations the number
of oscillator shells in the expansion is 12 for fermion wave functions,
and 20 for the meson fields. Of course 
for nuclei at the drip-lines, solutions in configurational representation
might not provide an accurate description of properties that
crucially depend on the spatial extension of nucleon densities, 
as for example nuclear radii. In the present study we 
are particularly interested in the deformed N=28 region: shell
effects and shape coexistence phenomena. For a description of 
general trends coordinate space solutions are not essential, 
solutions in the oscillator basis should present a pertinent
approximation.


\section{Deformed N=28 region: shell effects and shape coexistence}

The region of neutron-rich $N\approx 28$ nuclei presents many 
interesting phenomena: the average nucleonic potential is
modified, shell effects are suppressed, large quadrupole
deformations are observed as well as shape coexistence, 
isovector quadrupole deformations are predicted 
at drip-lines. The detailed knowledge of the microscopic structure of 
these nuclei is also essential for a correct description of the
nucleosynthesis of the heavy Ca, Ti and Cr isotopes. In the present
application of the RHB theory we study the structure of exotic neutron
rich-nuclei with $12\leq Z \leq 20$, and in particular the light $N=28$
nuclei. We are interested in the influence of the spherical shell 
$N=28$ on the structure of nuclei below $^{48}$Ca, in deformation 
effects that result from the $1f 7/2 \rightarrow fp$ core breaking, and
in shape coexistence phenomena predicted for these $\gamma$-soft nuclei.

The structure of exotic neutron-rich nuclei with $N\approx 28$ has 
been extensively studied by Werner et al.~\cite{Wer.94,Wer.96} in the
framework of the self-consistent mean-field theory: Skyrme Hartree-Fock
model and relativistic mean-field approach. Skyrme - HF calculations 
were performed by discretizing the energy functional on a three-dimensional 
Cartesian spline collocation lattice. The Skyrme interaction
SIII~\cite{BFG.75} was used. 
For the RMF model the Dirac equation for the baryons 
and the Klein-Gordon equations for the meson fields were solved using the
basis expansion methods. The NL-SH parameter set~\cite{SNR.93} was used for 
the mean-field Lagrangian. In both models the pairing correlations 
were included using the BCS formalism in the constant gap approximation
with a strongly reduced pairing interaction. HF and RMF results were
compared with the predictions of the finite-range droplet (FRDM) and
the extended Thomas-Fermi with Strutinsky integral (ETFSI) mass models,
as well as with available experimental data. In particular, the onset
of deformations around $N=28$, the stability of the heaviest Si, S, 
Ar and Ca isotopes, and the isovector deformations were investigated. 
Calculations confirmed strong deformation effects caused by the
the $1f 7/2 \rightarrow fp$ core breaking, especially in the RMF model.
Almost all deformed nuclei were found to be $\gamma$-soft, with the 
deformation  dependence of the single-particle spectrum favoring
prolate shapes for $Z=16$ and oblate for $Z=18$. The results obtained
with the two models, HF and RMF, were found to be similar, although 
also important differences were calculated. For example, the equilibrium
shape of the $N=28$ nucleus $^{44}$S. RMF calculations predict a well
deformed prolate ground state with $\beta_2 = 0.31$ for this nucleus, 
while according to the HF model, $^{44}$S is $\gamma$-soft with a small
quadrupole deformation $\beta_2 \approx 0.13$. The FRDM and ETFSI models
predict spherical and oblate ($\beta_2 = -0.26$) ground states, respectively.
Another detailed analysis of the ground-state properties of nuclei in the 
light mass region $10 \leq Z \leq 22$ in the framework of the RMF approach 
is reported in Ref.~\cite{LFS.98}. The calculations were also performed in
an axially deformed configuration using the NL-SH effective interaction. 
Pairing was included in the BCS approximation, but a somewhat different 
prescription was used to calculate the neutron and proton pairing gaps 
in regions where nuclear masses were not known~\cite{MN.92}. Nuclei at 
the stability line were considered, as well as those close to the proton 
and neutron drip-lines. Results of calculations were compared with
available empirical data and predictions of mass models. A very interesting 
result is the predicted ground-state deformation for $^{44}$S: oblate 
($\beta_2 = -0.2$), with a highly prolate shape ($\beta_2 = 0.38$) about 
30 keV above the ground state. This result is at variance with the 
calculations of Refs.~\cite{Wer.94,Wer.96}, although in both cases 
the same effective interaction was used for the mean-field Lagrangian.
The pairing gaps used in the two calculations however were different. 
Therefore it appears that, for nuclei which are very $\gamma$-soft,
the pairing interaction
determines the shape transition to the deformed intruder configurations.

Results of Skyrme - HF and RMF calculations clearly indicate that mean-field
and pairing correlations should be described in a unified self-consistent
framework: the nonrelativistic Hartree-Fock Bogoliubov theory~\cite{RS.80} 
or the relativistic Hartree-Bogoliubov model that was described in the
previous section. Only fully self-consistent HFB or RHB models 
correctly describe the virtual scattering of nucleonic pairs from 
bound states to the positive energy particle continuum~\cite{DNW.96}.
The correct representation of pairing correlations is
an essential ingredient for microscopic models of the 
structure of neutron-rich nuclei.

The input parameters for our model are the coupling constants and 
masses for the effective mean-field Lagrangian, and the 
effective interaction in the pairing channel. In the present
calculation we use the NL3 effective interaction~\cite{LKR.97}
for the RMF Lagrangian. We have used the NL3 force in most 
of the applications of RHB theory to spherical nuclei:
in the analysis of light neutron-rich nuclei in 
Refs. \cite{PVL.97,LVP.98,LVR.97}, in the study of ground-state
properties of Ni and Sn isotopes \cite{LVR.98}, and in the 
description of proton-rich nuclei with $14\leq Z \leq 28$ 
and $N = 18, 20, 22$.
Properties calculated with NL3 indicate that this is probably
the best RMF effective interaction so far, both for nuclei
at and away from the line of $\beta$-stability. NL3 has also 
been used in calculations of binding energies and deformation 
parameters of rare-earth nuclei~\cite{LR.98}. For the pairing 
field we employ the pairing part of the 
Gogny interaction, with the parameter set D1S \cite{BGG.84}. 

In Fig. \ref{figA} the two-neutron separation 
energies are plotted
\begin{equation}
S_{2n}(Z,N) = B_n(Z,N) - B_n(Z,N-2)
\label{sep}
\end{equation}
for the even-even nuclei $12\leq Z \leq 24$ and $24\leq N \leq 32$.
The values that correspond to the self-consistent RHB ground-states
(symbols connected by lines) 
are compared with experimental data and extrapolated values from 
Ref.~\cite{AW.95} (filled symbols). Except for Mg and
Si, the nuclei that we consider are not at the drip-lines.
The theoretical values 
reproduce in detail the experimental separation energies,
except for $^{48}$Cr. In general we have found that RHB model 
binding energies are in very good agreement with experimental data 
when one of the shells (proton or neutron) is closed, or 
when valence protons and neutrons occupy different major 
shells (i.e. below and above $N$ and/or $Z=20$). 
The differences are more pronounced when both protons and neutrons 
occupy the same major shell, and especially 
for the $N=Z$ nuclei. For these nuclei additional
correlations should be taken into account, and
in particular proton-neutron pairing 
could have a strong influence on the masses. Proton-neutron 
short-range correlations are not included in our model.

In Fig. \ref{figB} the calculated neutron radii are displayed.
The model of course predicts an increase of the calculated 
radii with the number of neutrons. It is interesting however
that Cr, Ti, Ca, Ar, and to a certain extent S isotopes, display
very similar neutron radii in this region. Only for the two more
exotic chains of Si and Mg isotopes, i.e. more neutron-rich, a 
substantial increase of the calculated radii is observed.  
Very important for our present investigation is the effect 
that the $N = 28$ spherical shell closure produces on the 
neutron radii of Cr, Ti and Ca isotopes. On the other hand,
no effect of shell closure is observed for S, Si and Mg. This 
is already an indication that the $N = 28$ shell effects are 
suppressed in neutron-rich nuclei.

The predicted mass quadrupole deformations for the ground states 
of $N=28$ nuclei are shown in the upper panel of Fig. \ref{figC}.
We notice a staggering between prolate and oblate configurations, 
and this indicates that the potential is $\gamma$-soft. The absolute
values of the deformation decrease as we approach the $Z=20$ closed
shell. Starting with Ca, the $N=28$ nuclei are spherical in the ground
state. The calculated quadrupole deformations are in agreement
with theoretical results reported by Werner et al.~\cite{Wer.94,Wer.96} 
(prolate for $Z=16$, oblate for $Z=18$), and with available experimental 
data: $| \beta_2 | = 0.258 (36)$ for $^{44}$S ~\cite{Glas.97,Glas.98},
and $| \beta_2 | = 0.176 (17)$ for $^{46}$Ar~\cite{Sch.96}.
Experimental data (energies of $2_1^+$ states and 
$B(E2; 0_{g.s.}^+ \rightarrow 2_1^+)$ values) do not determine the 
sign of deformation, i.e. do not differentiate between prolate and 
oblate shapes. In the lower panel of
Fig. \ref{figC} we display the average values of the neutron 
pairing gaps for occupied canonical states
\begin{equation}
< \Delta_N > = {{\sum_{[\Omega \pi]} \Delta_{[\Omega \pi]} 
v_{[\Omega \pi]}^2}\over
                  {\sum_{[\Omega \pi]}  v_{[\Omega \pi]}^2}},
\label{ang}
\end{equation}
where $v_{[\Omega \pi]}^2$ are the occupation probabilities, and
$\Delta_{[\Omega \pi]}$ are the diagonal matrix elements
of the pairing part of the RHB single-nucleon Hamiltonian in the 
canonical basis.
$< \Delta_N >$ provides an excellent quantitative measure
of pairing correlations. The calculated values of  
$< \Delta_N > \approx 2$ MeV correspond to those found in 
open-shell Ni and Sn isotopes~\cite{LVR.98}. The spherical
shell closure $N=28$ is strongly suppressed for nuclei
with $Z\leq 18$, and only for $Z\geq 20$ neutron pairing 
correlations vanish. It is also interesting to notice how
in the self-consistent RHB model  
the calculated gaps reflect the staggering of ground-state
quadrupole deformations. 

For $^{44}$S in 
Fig. \ref{figD} we plot the neutron canonical pairing gaps 
$\Delta_{[\Omega \pi]}$ 
as function of canonical single-particle energies. The gaps 
are displayed for canonical states that 
correspond to the self-consistent ground state. The dashed line
denotes the position of the Fermi energy.
The pairing gaps are more or less constant for deeply bound states and 
display a sharp decrease at the Fermi surface.
This is related to the volume character of the Gogny interaction in the 
pairing channel. The values of the $\Delta_{[\Omega \pi]}$ are around 2.2 MeV in
the volume, and the average value at the Fermi surface is around 1.8 MeV. 

The results of Skyrme+HF and RMF calculations of Refs.~ \cite{Wer.94,Wer.96}
have shown that neutron-rich Si, S and Ar isotopes can be considered as
$\gamma$-soft, with deformations depending on subtle interplay between 
the deformed gaps $Z=16$ and $18$, and the spherical gap at $N=28$.
Because of cross-shell excitations to the $2p_{3/2}$, $2p_{1/2}$ and 
$1f_{5/2}$ shells, the $N=28$ gap appears to be broken in most cases.
In the RMF analysis of Ref.~\cite{LFS.98} a careful study of the 
phenomenon of shape coexistence was performed for nuclei in this region.
It was shown that several Si and S isotopes exhibit shape coexistence:
two minima with different deformations occur in the binding energy. 
The energy difference between the two minima is of the order of 
few hundred keV. For $^{44}$S this difference was found to be only
30 keV. In the fully microscopic and self-consistent RHB model, we 
have the possibility to analyze in detail the single-neutron levels
and to study the formation of minima in the binding energy. In 
Figs.~\ref{figE}-\ref{figG} we display the single-neutron levels
in the canonical basis for the $N=28$ nuclei $^{42}$Si, $^{44}$S,
and $^{46}$Ar, respectively. The single-neutron eigenstates of the
density matrix result from constrained RHB calculations performed
by imposing a quadratic constraint on the quadrupole moment. 
The canonical states are plotted as function of the quadrupole 
deformation, and the dotted curve denotes the position of the 
Fermi level. In the insert we plot the corresponding total binding 
energy curve as function of the quadrupole moment. For $^{42}$Si
the binding energy displays a deep oblate minimum ($\beta_2 \approx -0.4$).
The second, prolate minimum is found at an excitation energy of 
$\approx 1.5$ MeV. Shape coexistence is more pronounced for $^{44}$S.
The ground state is prolate deformed, the calculated deformation 
in excellent agreement with experimental data~\cite{Glas.97,Glas.98}.
The oblate minimum is found only $\approx 200$ keV above the ground state. 
Finally, for the nucleus $^{46}$Ar we find a very flat energy 
surface on the oblate side. The deformation of the ground-state
oblate minimum agrees with experimental
data~\cite{Sch.96}, the spherical state is only few keV higher. 
It is also interesting to observe how the spherical gap between
the $1f_{7/2}$ orbital and the $2p_{3/2}$, $2p_{1/2}$ orbitals
varies with proton number. While the gap is really strongly reduced 
for $^{42}$Si and $^{44}$S, in the $Z=18$ isotone $^{46}$Ar 
the spherical gap is $\approx 4$ MeV. Of course from $^{48}$Ca  
the $N=28$ nuclei become spherical. Therefore the single-neutron 
canonical states in Figs.~\ref{figE}-\ref{figG} clearly display 
the disappearance of the spherical $N=28$ shell closure for 
neutron-rich nuclei below $Z=18$.

In order to illustrate the importance of the correct description of pairing
correlations, in Figs.~\ref{figH}-\ref{figK} we compare results of 
fully self-consistent RHB calculations with those obtained by a simplified 
RMF+BCS approach that was also used in Refs.~\cite{Wer.94,Wer.96,LFS.98}.
Binding energies, neutron and proton rms radii, and ground-state quadrupole
deformations are compared for chains of Mg, Si, S and Ar isotopes. In both
models the NL3 effective interaction has been used for the mean-field 
Lagrangian. The D1S Gogny interaction in the pairing 
channel of the RHB calculation, the constant gap approximation in the 
BCS scheme. Since for many of these nuclei the experimental odd-even 
mass differences are not known, the proton and neutron gaps 
are calculated following the prescription of M\" oller and Nix~\cite{MN.92}
\begin{equation}
\Delta_n = {4.8 \over {N^{1/3}} }~~~~~~~~~~\Delta_p = {4.8 \over {Z^{1/3}} }
\end{equation}
From Figs.~\ref{figH}-\ref{figK} we notice that, although the calculated
binding energies are almost identical with a possible exception for Mg at
the drip-line, the other quantities display significant differences. RMF+BCS
systematically predicts larger rms radii, especially for protons. For neutron
radii RMF+BCS calculations produce interesting kinks at $N=28$ for Mg, Si and
Ar, which suggest shell closure. In the case of Ar isotopes, a
strong shift is also calculated for the proton rms radius at N=28. 
These discontinuities are not found 
in RHB results, which display a uniform increase of radii with neutron number. 
The two models predict similar ground state quadrupole deformations, 
except for the very $\gamma$-soft Ar isotopes. For $^{42,44}$Ar the differences
are rather pronounced, the ground-states of $^{44}$Ar is prolate in RMF+BCS,   
slightly oblate in RHB calculations. 

In conclusion, in the present work we have performed a detailed 
analysis of the deformed $N=28$ region of neutron-rich nuclei in the 
framework of the relativistic Hartree-Bogoliubov theory with finite
range pairing interaction. This is the first application of the RHB 
model to the structure of deformed nuclei. In particular, we 
have investigated the suppression of the spherical $N=28$ shell
gap for neutron-rich nuclei, and the related phenomenon of shape 
coexistence. The NL3 effective interaction has been used for the 
mean-field Lagrangian, and pairing correlations have been described 
by the pairing part of the finite range Gogny interaction D1S. 
Two-neutron separation energies, rms radii, and ground-state 
quadrupole deformations have been compared with available 
experimental data and results of previous Skyrme Hartre-Fock
and relativistic mean-field + BCS calculations. We have shown 
that the present version of the RHB model produces results in excellent
agreement with experimental data, both for binding energies and 
quadrupole deformations. Our results confirm the strong deformations
caused by $1f7/2 \rightarrow fp$ neutron excitations. Constrained 
RHB calculations have been used to construct the binding energy curves as
function of quadrupole deformation. Pronounced shape coexistence 
is found to be a characteristic of neutron-rich nuclei in this region.
$N\approx 28$ nuclei become extremely $\gamma$-soft just before the 
closure of the $Z=20$ shell. For $Z\leq 16$ single neutron spectra in the
canonical basis of RHB display a strong reduction of the gap between the 
$1f_{7/2}$ orbital and the $2p_{3/2}$, $2p_{1/2}$ levels. The suppression
of the spherical $N=28$ shell closure favors deformed ground states.
The importance of the full RHB description of pairing correlations has 
been illustrated by a comparison with ground-state properties of
Mg, Si, S and Ar isotopes calculated in the RMF+BCS scheme with 
constant gap approximation.
  
\bigskip
\begin{center}
{\bf ACKNOWLEDGMENTS}
\end{center}

This work has been supported in part by the 
Bundesministerium f\"ur Bildung und Forschung under
project 06 TM 875, and by the
Bulgarian National Foundation for
Scientific Research under project $\Phi $-527.

\newpage

\centerline{\bf Figure Captions}
\bigskip

\begin{figure}
\caption{ Two-neutron separation energies in the $N \approx 28$ 
region calculated in the RHB model and compared with 
experimental data (filled symbols) 
from the compilation of G. Audi and A. H. Wapstra.}
\label{figA}
\end{figure}

\begin{figure}
\caption{ Calculated neutron rms radii for the ground-states 
of nuclei in the $N \approx 28$ region: $12 \leq Z\leq 24$ 
and $24 \leq N\leq 32$.} 
\label{figB}
\end{figure}

\begin{figure}
\caption{ Self-consistent RHB quadrupole deformations for  
ground-states of the $N = 28$ isotones  (upper panel). 
Average neutron pairing gaps $< \Delta_N >$ 
as function of proton number (lower panel).}
\label{figC}
\end{figure}

\begin{figure}
\caption{ Average values of the neutron canonical pairing gaps 
as function of canonical single-particle energies for states that 
correspond to the self-consistent ground state of $^{44}$S. The NL3 
parametrization has been used for the mean-field
Lagrangian, and the parameter set D1S for the 
pairing interaction.}
\label{figD}
\end{figure}

\begin{figure}
\caption{ The neutron single-particle levels for $^{42}$Si as
function of the quadrupole deformation. The energies in the 
canonical basis correspond to ground-state RHB solutions 
with constrained quadrupole deformation. The dotted line 
denotes the neutron Fermi level. In the insert we display 
the corresponding total binding energy curve.} 
\label{figE}
\end{figure}

\begin{figure}
\caption{ The same as in Fig. 5, but for $^{44}$S.}   
\label{figF}
\end{figure}

\begin{figure}
\caption{ The same as in Fig. 5, but for $^{46}$Ar.}   
\label{figG}
\end{figure}

\begin{figure}
\caption{ Comparison of ground-state properties of neutron-rich 
Mg isotopes, calculated with the relativistic Hartree-Bogoliubov
model (NL3 plus D1S Gogny in the pairing channel) and with
the relativistic mean-field plus BCS model (NL3 plus constant-gap
approximation in the pairing channel). Total binding energies,
neutron and proton rms radii, and ground-state quadrupole 
deformations are compared.}
\label{figH}
\end{figure}

\begin{figure}
\caption{ The same as in Fig. 8, but for the Si isotopes.}   
\label{figI}
\end{figure}

\begin{figure}
\caption{ The same as in Fig. 8, but for the S isotopes.}   
\label{figJ}
\end{figure}

\begin{figure}
\caption{ The same as in Fig. 8, but for the Ar isotopes.}   
\label{figK}
\end{figure}

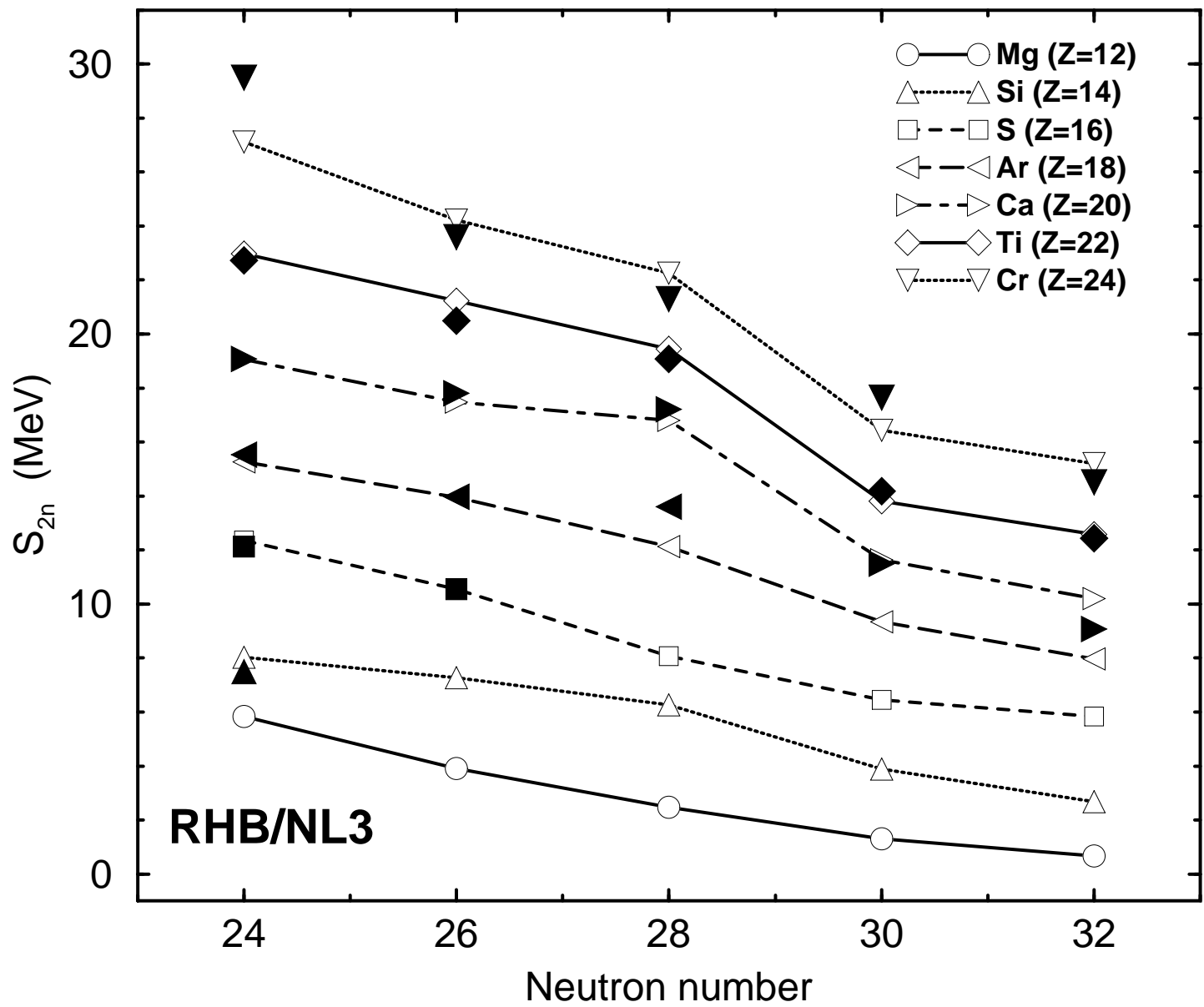
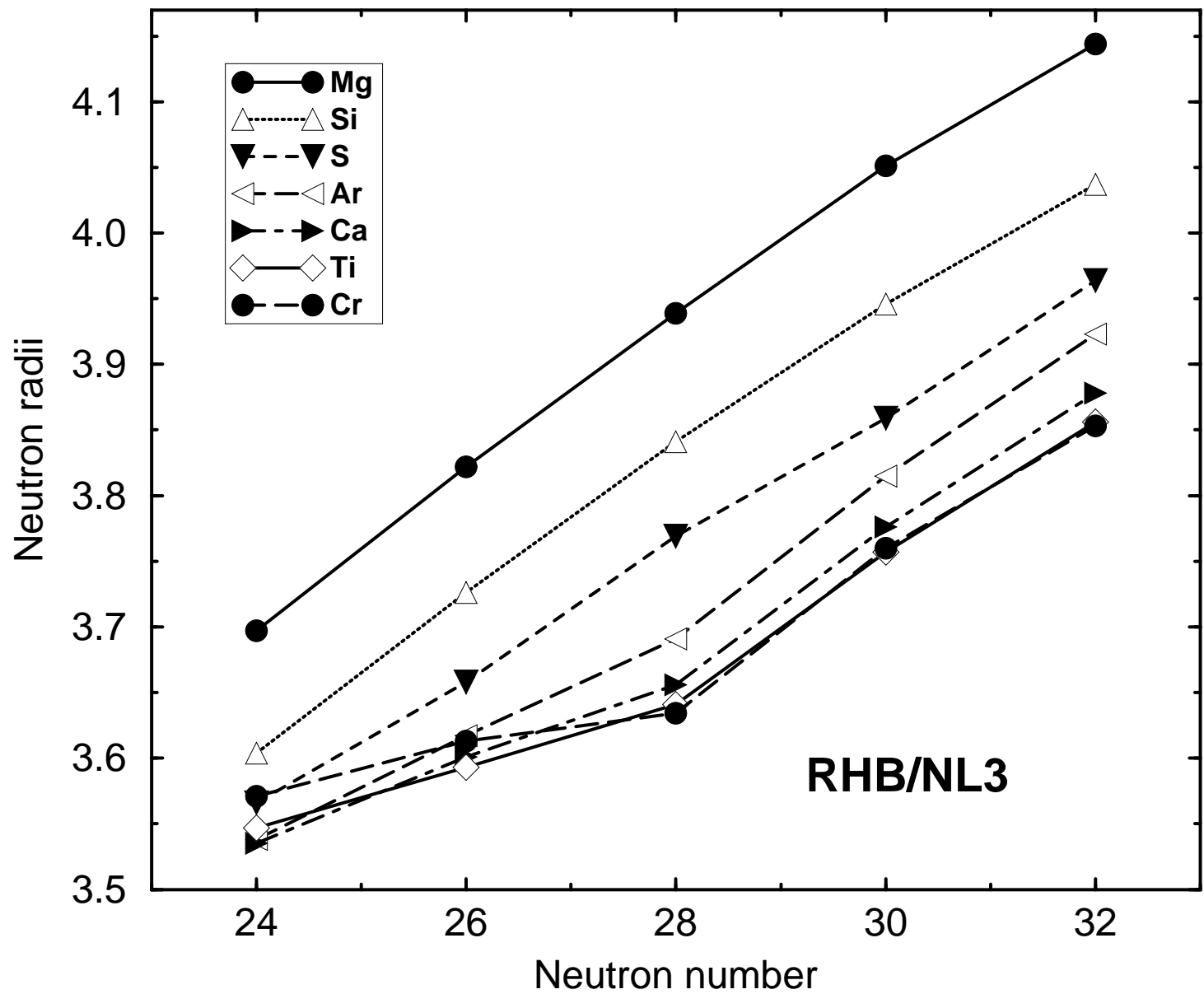
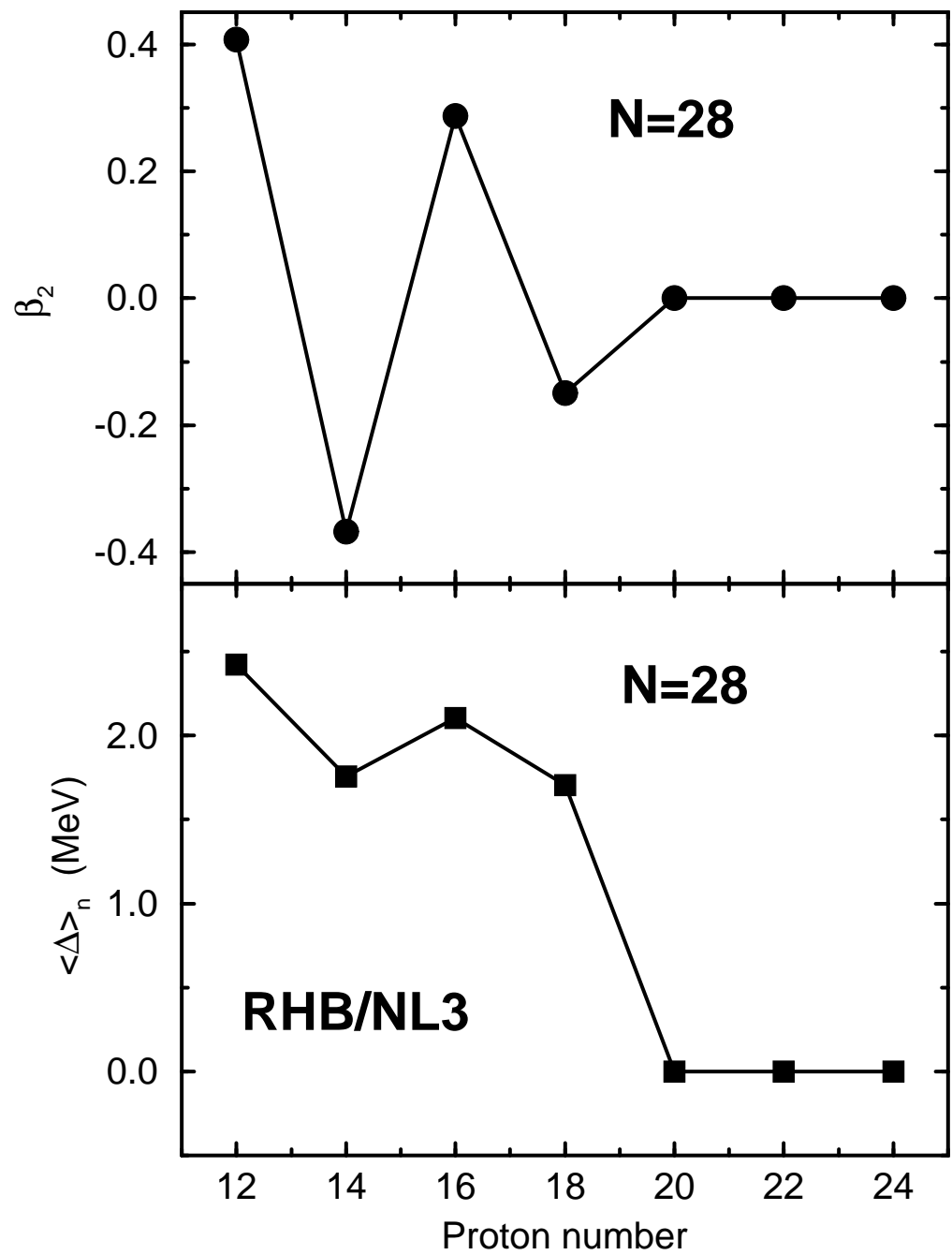
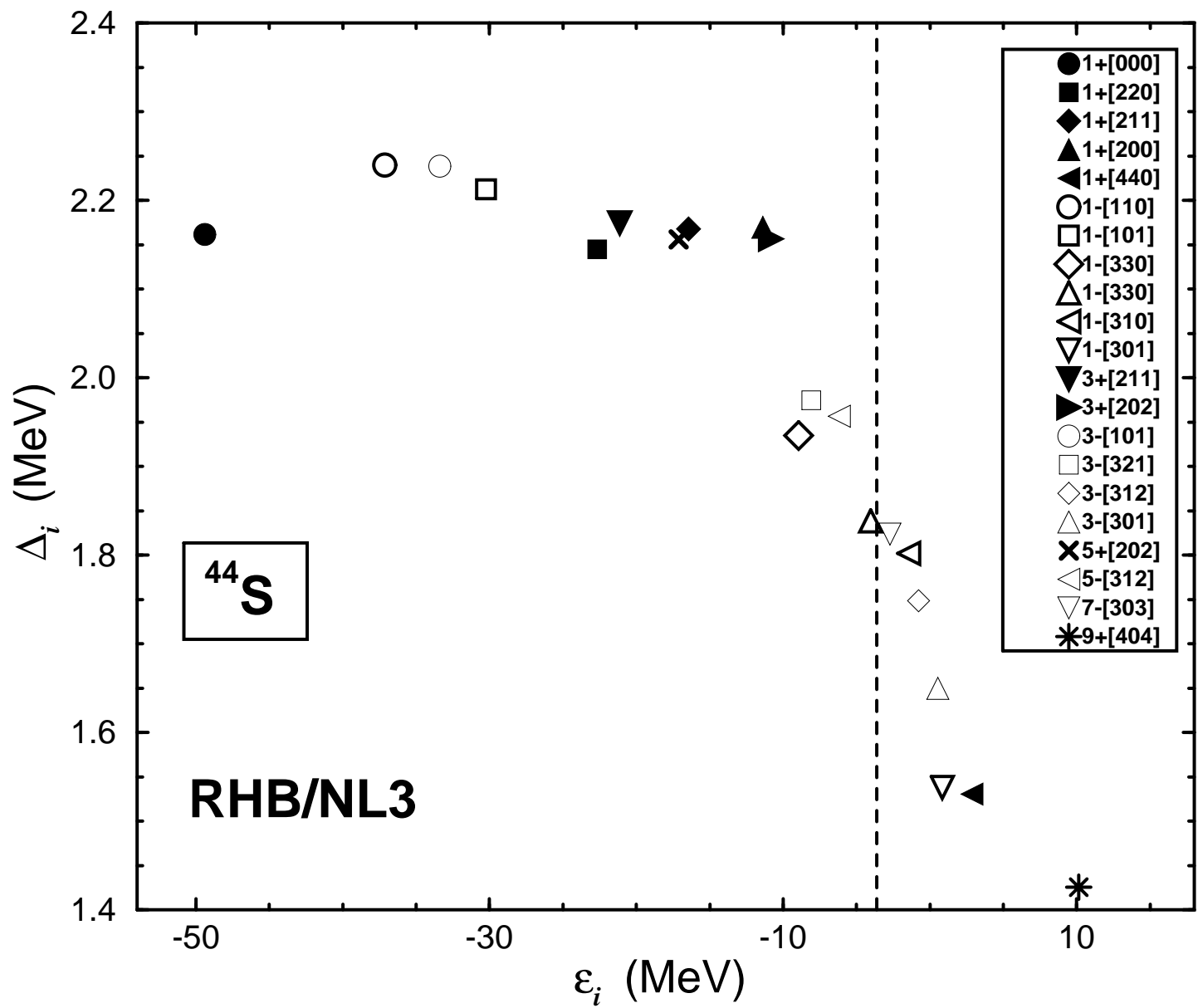
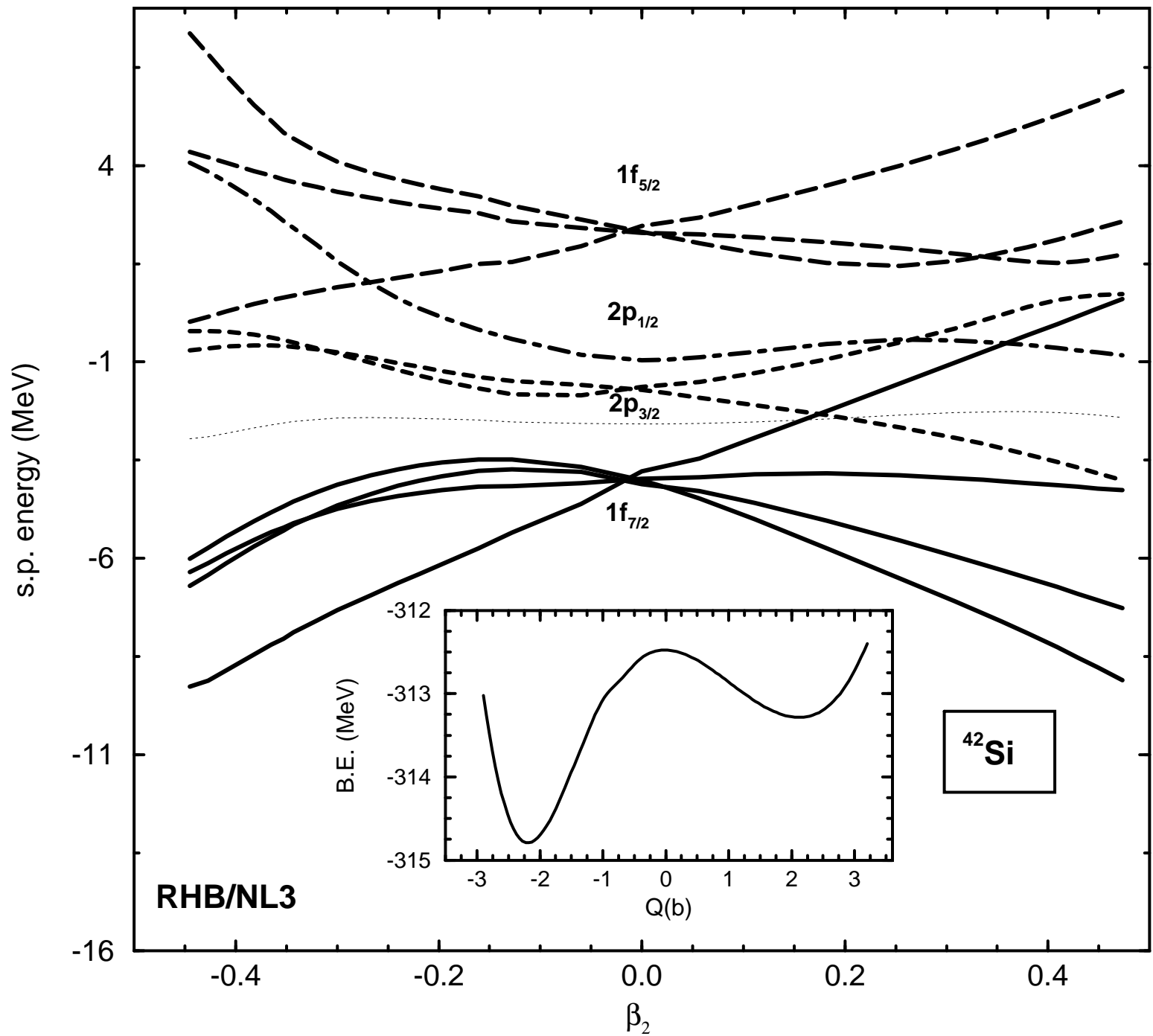
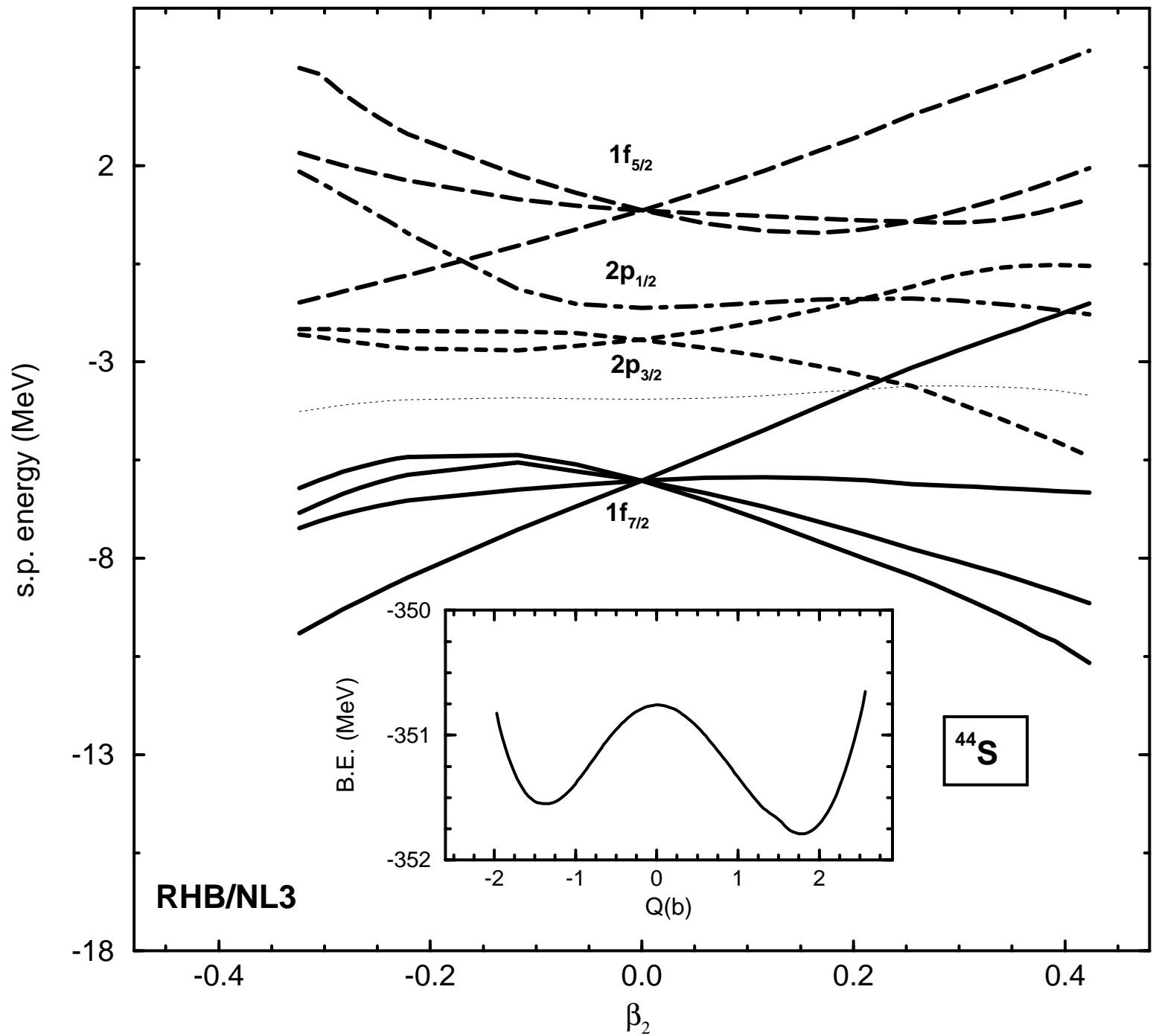
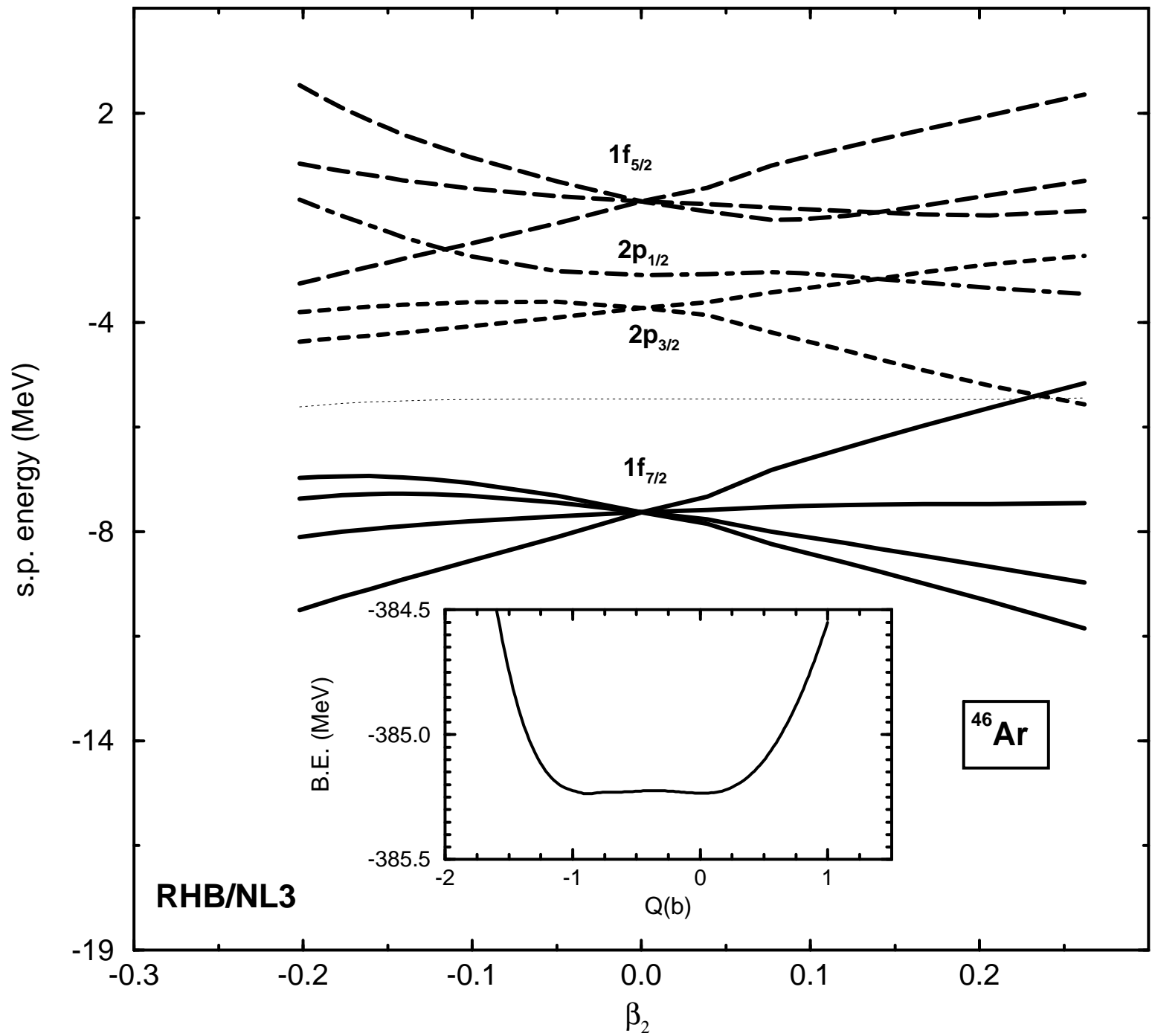
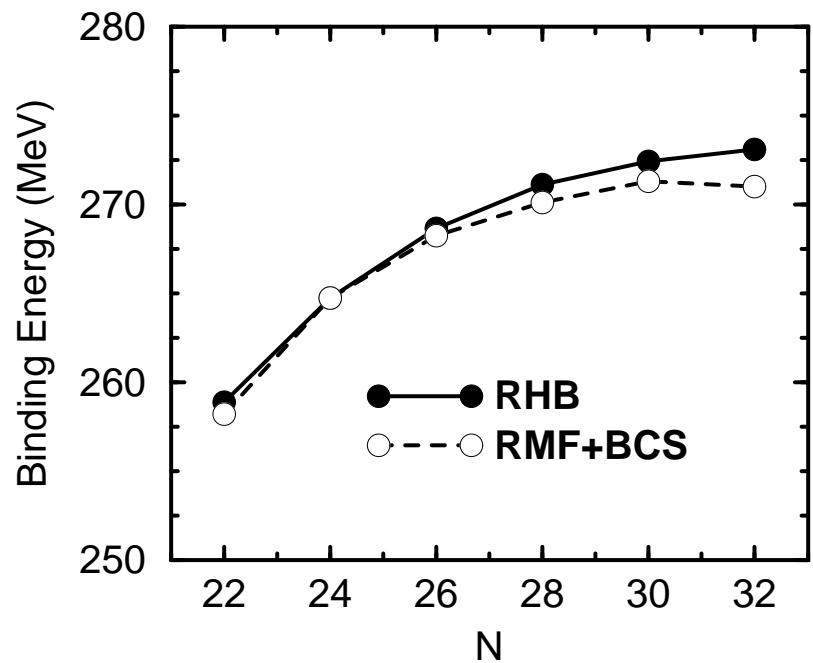
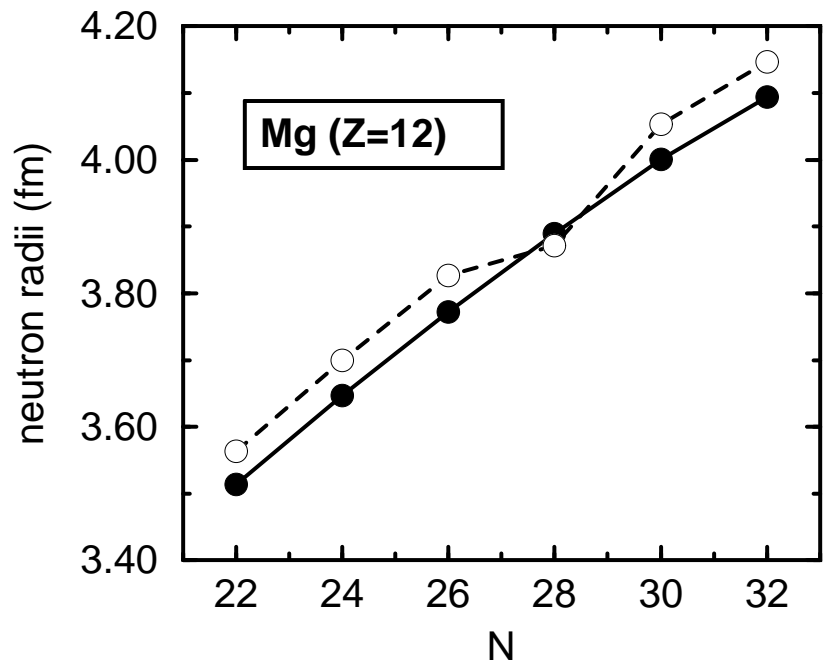
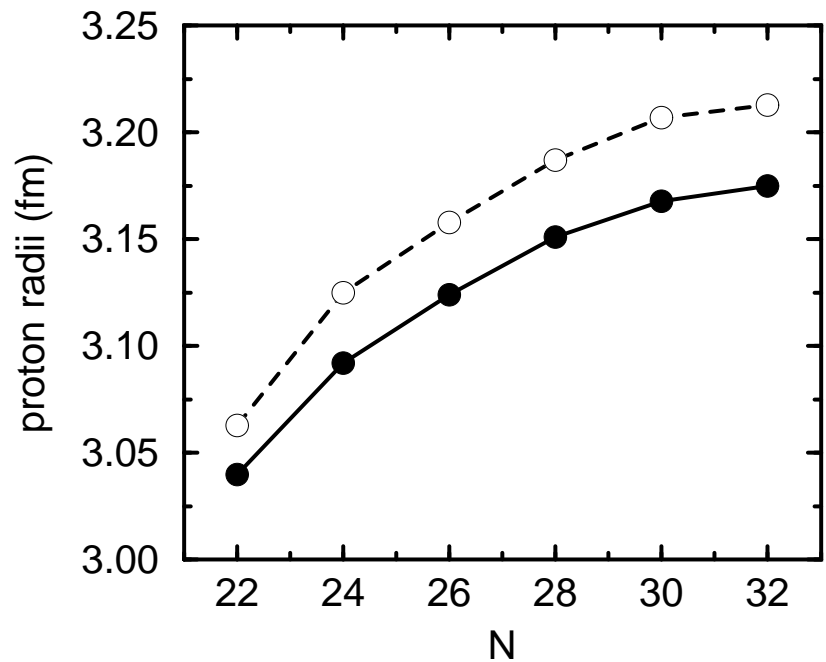
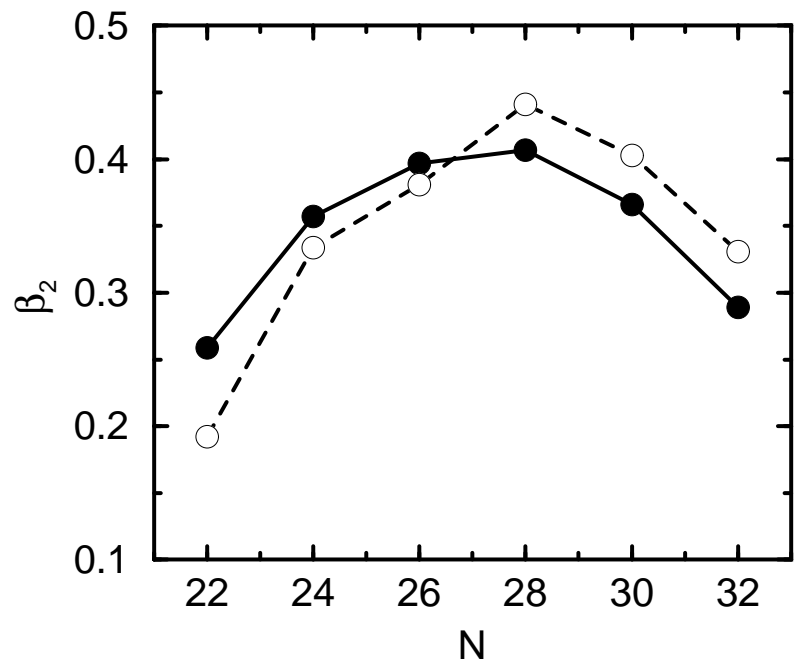
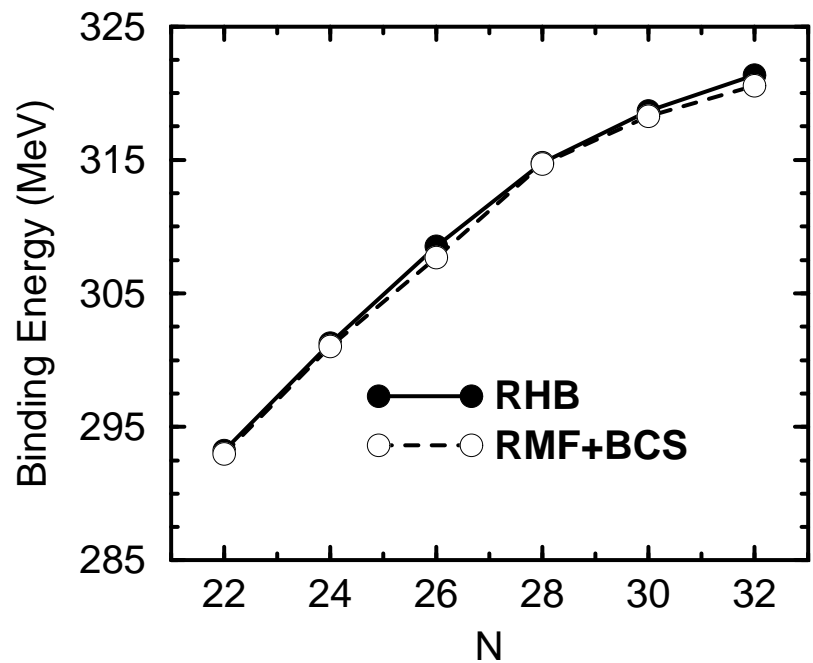
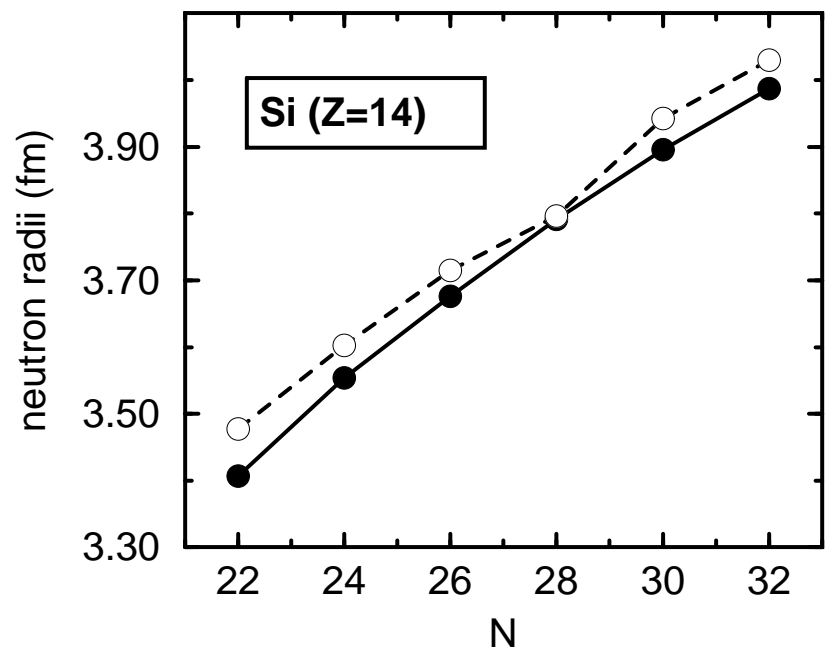
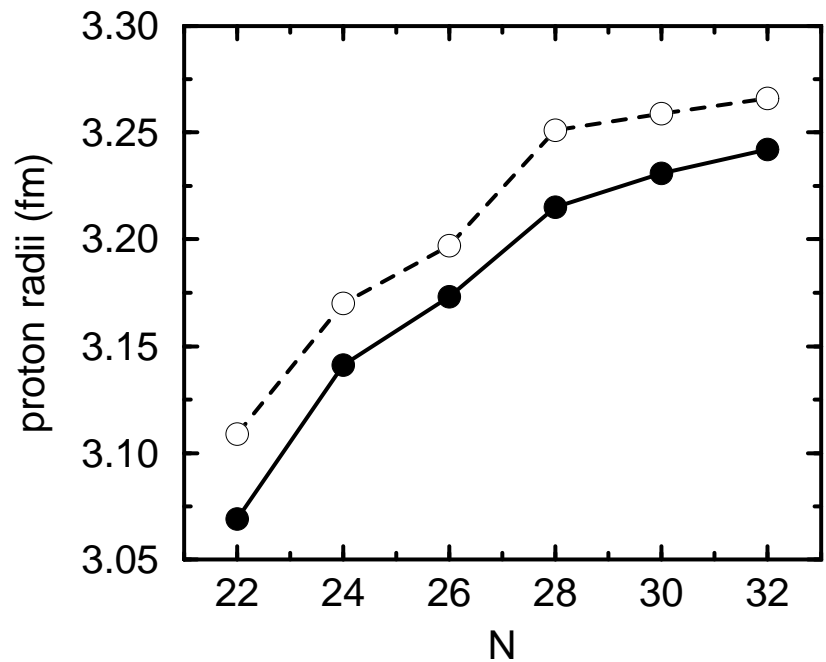
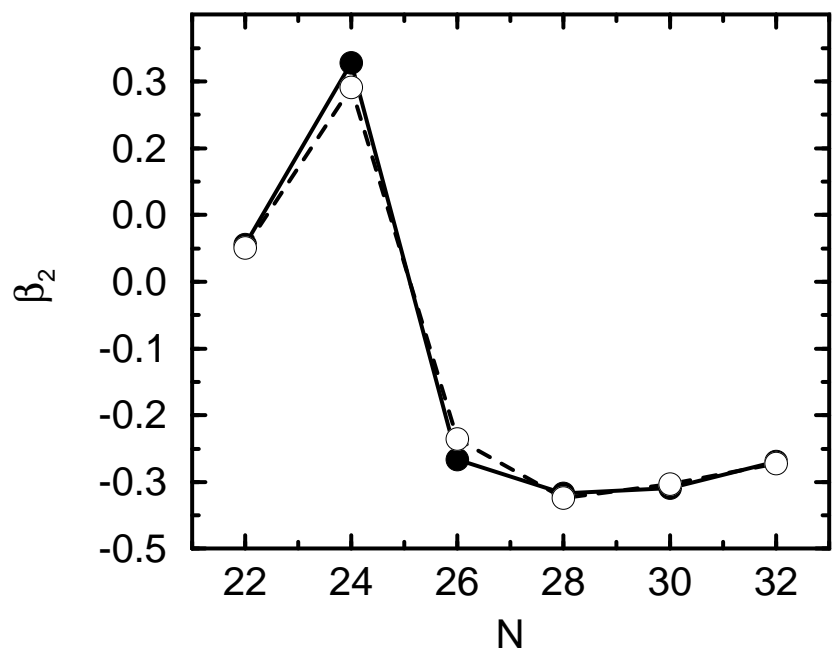
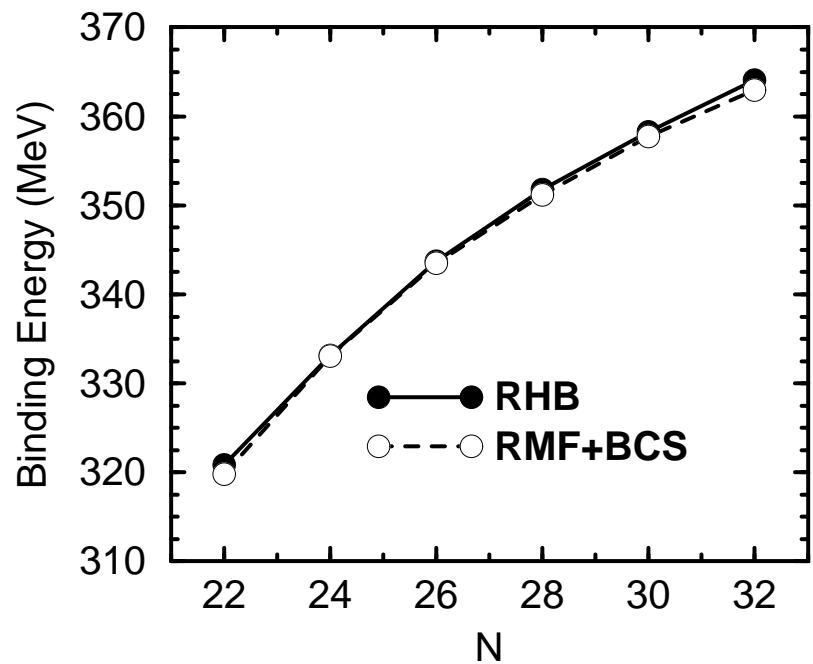
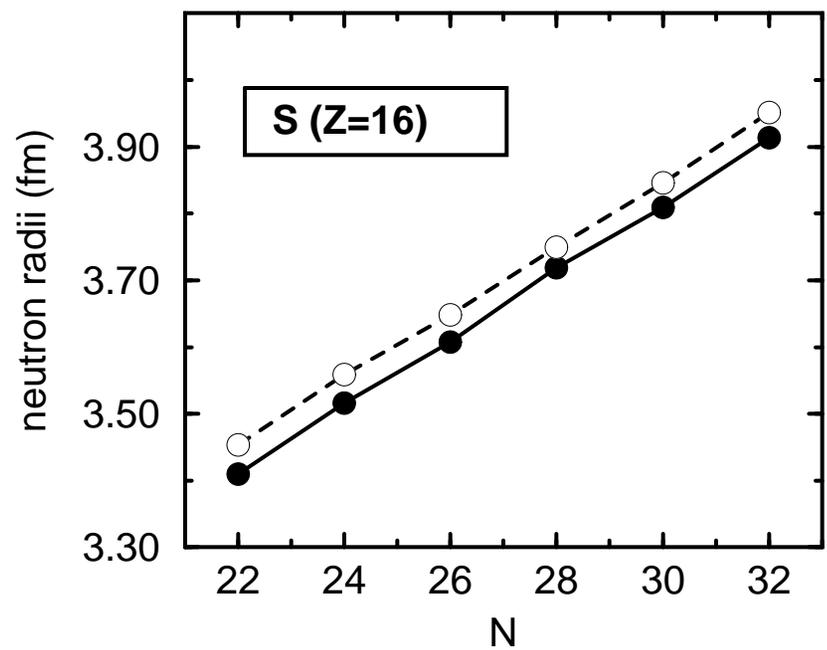
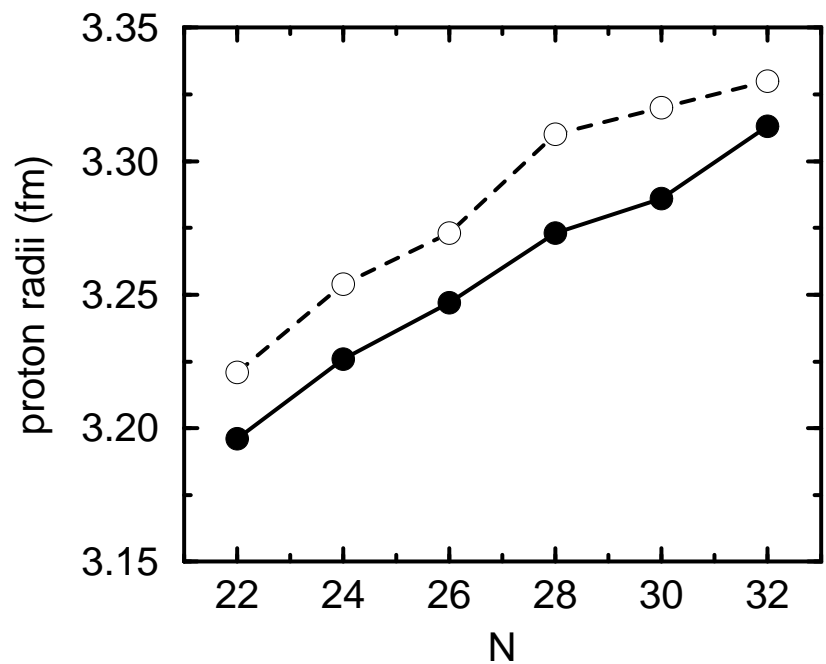
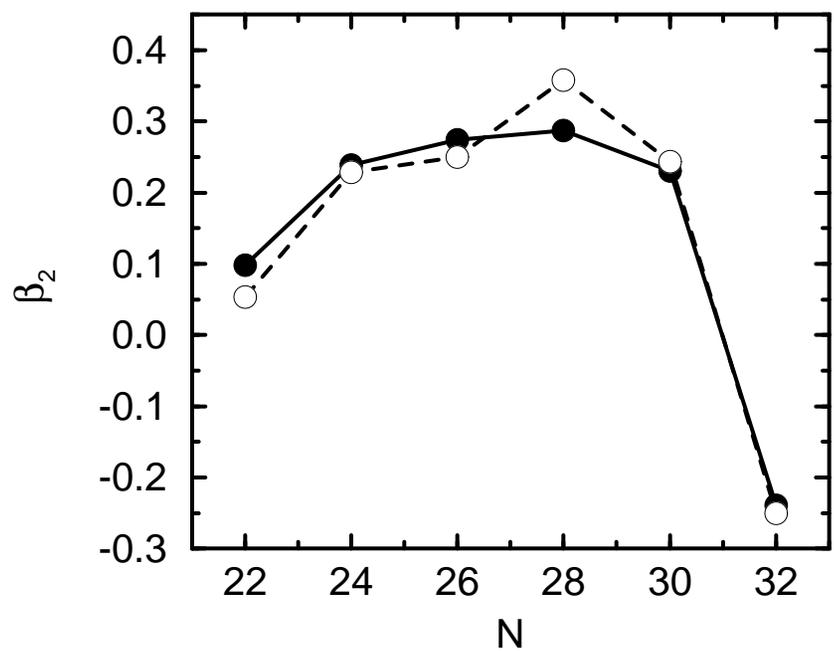
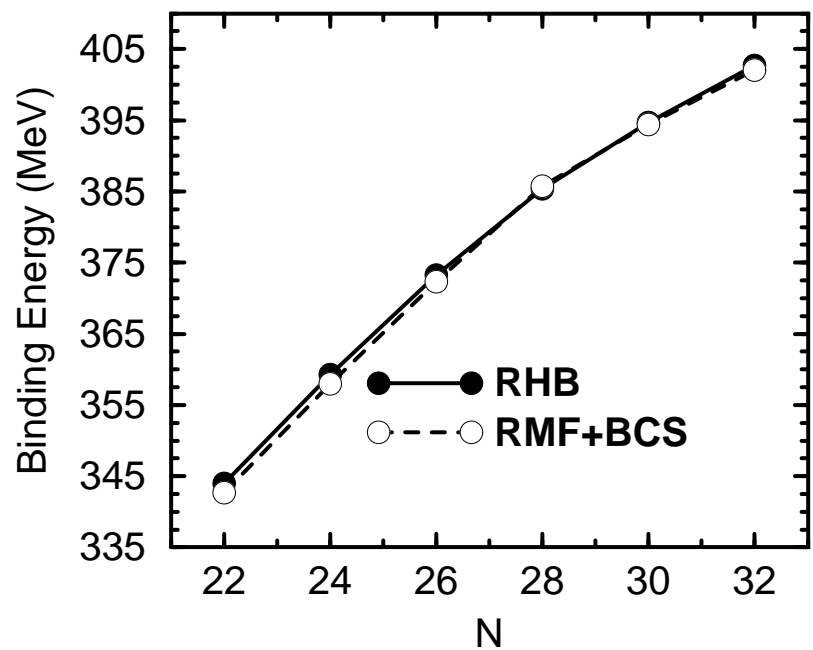
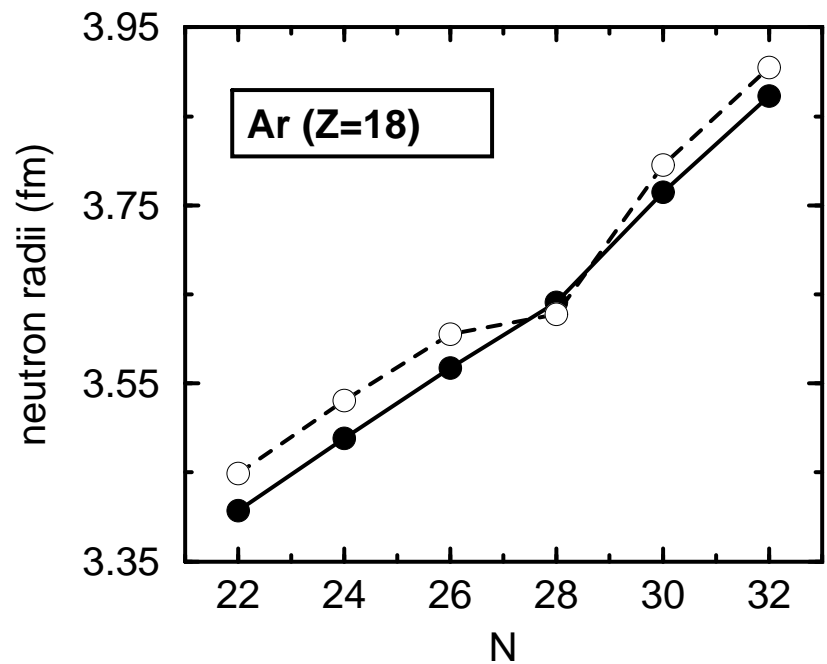
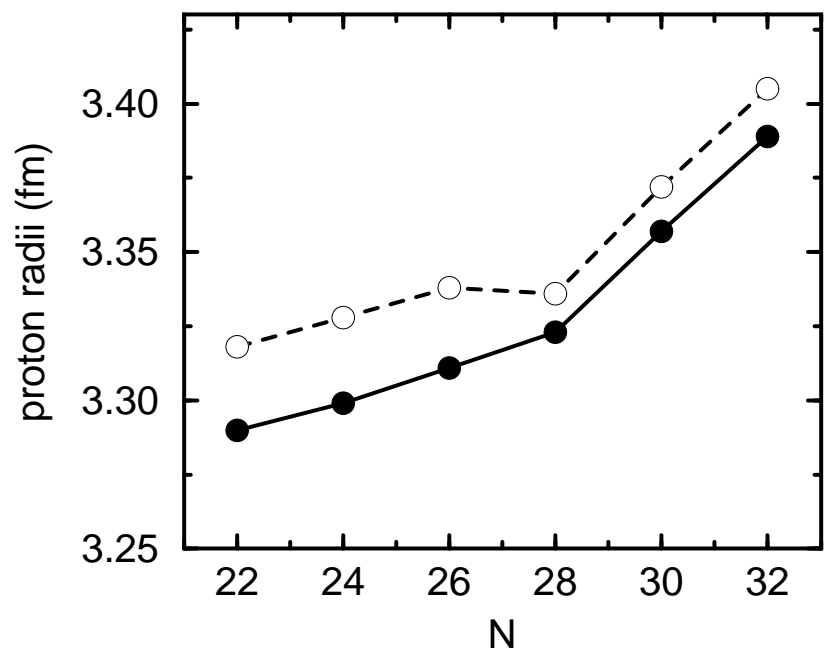
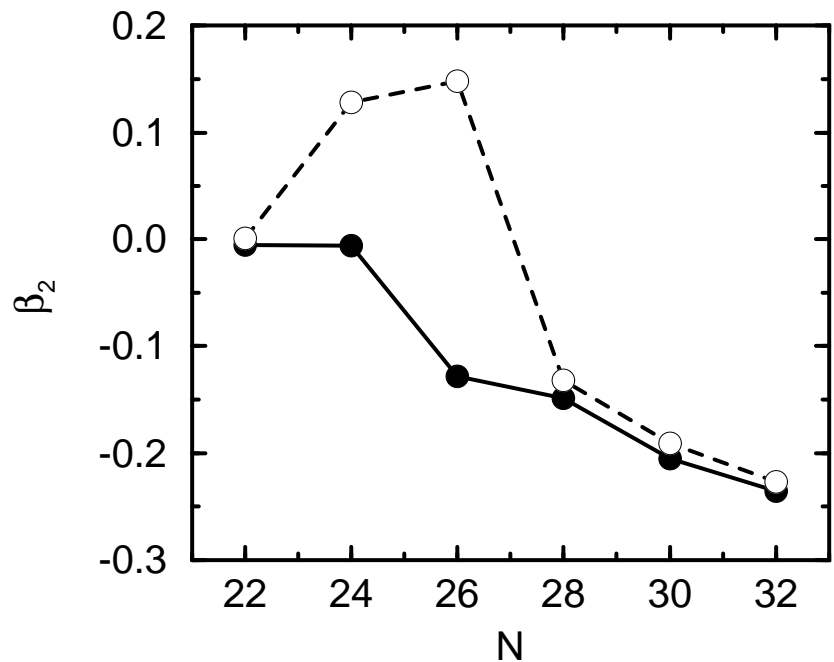
\end{document}